\documentstyle[preprint,aps]{revtex}
\begin{document}
\title{SUPERSYMMETRIC CLASSICAL MECHANICS: FREE CASE}
\author{R. de Lima Rodrigues$^{(a)}$\thanks{E-mail: rafael@cfp.ufpb.br},
W. Pires de Almeida$^{(b)}$
and I. Fonseca Neto$^{(c)}$\\
 {}$^{(a)}$ Centro Brasileiro de Pesquisas F\'\i sicas\\
Rua Dr. Xavier Sigaud, 150\\
CEP 22290-180, Rio de Janeiro-RJ, Brazil\\
{}$^{(a,b)}$ Departamento de Ci\^{e}ncias Exatas e da\\
Natureza, Universidade Federal da Para\'{\i}ba\\
Campus V, Cajazeiras, PB, 58.900-000, Brazil\\
{}$^{(c)}$ Departamento de F\'{\i}sica,
Universidade Federal da Para\'{\i}ba\\
Campus II, Campina Grande, PB, 58.108-900, Brazil}

\maketitle

\vspace{1cm}

\centerline{\bf{ABSTRACT}}

We present a review work on Supersymmetric Classical Mechanics in
the context of a Lagrangian formalism, with $N=1-$supersymmetry. We show
that the $N=1$ supersymmetry does not allow the introduction of a
potential energy term
depending on a single commuting supercoordinate, $\phi (t;\Theta )$.

\pacs
\newpage

\section{INTRODUCTION}
\label{sec:level1}

Supersymmetry (SUSY) in classical mechanics (CM) \cite{Galv80,Salo82} in a
non-relativistic scenario is investigated. SUSY first appeared in
relativistic theories in terms of bosonic and fermionic fields\footnote{A
bosonic field (associated with particles of integral or null spin) is one
particular case obeying the Bose-Einstein statistic and a fermionic field
(associated to particles with semi-integral spin) is that obey the
Fermi-Dirac statistic.}, and the possibility was early observed that it can
accomodate a Grand-Unified Theory (GUT) for the four basic interactions of
Nature (strong, weak, electromagnetic and gravitational). However, after a
considerable  number of works investigating SUSY in this context,
confirmation of SUSY as high-energy unification theory is missing.
Furthermore, there exist phenomenological applications of the $N=2$ SUSY
technique in quantum mechanics (QM) \cite{Ber84}. In the literature, there
exist four excellent review articles about SUSY in quantum mechanics
\cite{Gen}. Recently a general review on the SUSY QM algebra and the
procedure on like to build a SUSY Hamiltonian hierarchy in order of a
complete spectral resolution it is explicitly applied for the
P\"oschl-Teller I potential \cite{rafa01}.

We must say that, despite being introductory, this work is not a mere
scientific exposition. It is intended for students and teachers of science
and technology. The pre-requisites are differential and integral
calculus of two real variables functions and classical mechanics.
Recently, two excellent mini-courses were ministered of introduction
to the theory of fields with the aim of presenting the
fundamental basics of the theory of fields including the idea of SUSY
with emphasis on basic concepts and a pedagogical
introduction to weak scale supersymmetry phenomenology, in which
the reader may use for different  approaches and viewpoints include
\cite{Swieca93}.

Considering two ordinary real variables  $x$ and $y$, it is well known that
they obey the commutative property, $xy=yx$. However, if
$\tilde x$ and $\tilde y$ are real Grassmann variables, we
have: $\tilde x\tilde y = - \tilde y\tilde x \Rightarrow \tilde x^2
= -\tilde x^2 = 0$.

In this work, we will use a didactic approach about the transformations
in the superspace, showing the infinitesimal transformation laws of the
supercoordinate and of its components, denominated by even and odd
coordinates, in the unidimensional space-time {D=(0+1) = 1}. We will see
that by making an
infinitesimal variation in the even coordinate, we generate the odd
coordinate and vice-versa. This approach is done in this
work using the right derivative
rule. We will distinguish that property of supersymmetry in which the
action is invariant with the translation transformations in the
superspace ($\delta S=0$), noting that the same does not occur with the
Lagrangian ($\delta L\neq 0$).

In the construction of a SUSY theory with $N>1$, referred to as extended SUSY,
for each spatial commuting coordinate, representing the degrees of freedom
of the system, we associate one anticommuting variable,
which are known that Grassmannian variables.
However, we consider only the $N=1$ SUSY for a
non-relativistic point particle, which is described by the introduction of only one real Grassmannian variable $\Theta$, in the configuration space,
but all the dynamics are putted in the time $t$. In this case, we have two
degrees of freedom. The generalized anticommuting coordinate (odd magnitude)
will be represented by $\psi(t)$. The new real coordinate
defined in the superspace will be called supercoordinate. It will have the
following more general possible Taylor expantion:
$\phi(t;\Theta) = q(t) + i \Theta\Psi(t)$. Note that the first term is exactly the ordinary real
commuting coordinate $q(t)$ and, like the next term, must to be linear
in $\Theta$, because $\Theta^2 = 0$. In this case, the time dependent part
multiplying $\Theta$ is necessarely one Grassmannian variable $\Psi(t)$,
which need the introduction of $i$ for warranty that the supercoordinate
$\phi(t;\Theta)$ will be real\footnote{Like will be see below.}.

We would like to highlight to the readers who know field theory, but have
never seen the supersymmetrical formalism in the context of classical
mechanics that in this work we will present the ingredients for implenting
N=1 SUSY, namely, superspace, supertranslation,
supercoordinate, SUSY covariant derivative and super-action. Indeed,
the steps adopted in this approach are the same as used in the
supersymmetrizations of quantum field theories in the quadrimensional
space-time of the special relativity (D = (3+1), three position coordinates
and one temporal coordinate).

This work is organized as follow: in section II we construct
a finite supercoordinate transformation and the infinitesimal transformations
on the supercoordinate and  its components via the translaction
in the superspace. In section III, we investigate the superparticle
using the Lagrangian formalism in the superspace, noting the fact that the
N=1 SUSY does not allow the introduction of a potential term for only one
supercoordinate and we indicate the quantizating procedure. In section
IV, we present the conclusion.

\section{ TRANSLATIONS IN SUPERSPACE}
\label{sec:level2}

We will consider the N=1 supersymmetry i.e. SUSY with only one
anticommuting variable. Supersymmetry in classical mechanics unify the
even coordinate $q(t)$ and the odd cordinate $\Psi(t)$ in a superspace characterized
by the introduction of a Grassmannian variable, $\Theta$, not measurable
\cite{Galv80,Salo82,Berezin66}.

\begin{equation}
\label{E(l)}
\hbox{Superspace} \rightarrow (t;\Theta ), \qquad \Theta ^{2}=0,
\end{equation}
where $t$ and $\Theta$ act, respectively, like even and odd elements of the
Grassmann algebra.

The anticommuting coordinate, $\Theta $, will parametrize all points of
superspace, but all dynamics will be put in the time coordinate $t$. SUSY
in classical mechanics is generated by a translation transformation
in the superspace, viz.,

\begin{eqnarray}
\label{E2}
\Theta&\rightarrow& \Theta^{\prime}= \Theta + \epsilon , \Rightarrow
\delta\Theta = \Theta^{\prime} - \Theta = \epsilon \nonumber\\
t &\rightarrow& t^{\prime}= t + i\epsilon \Theta \Rightarrow
\delta t = t^{\prime} - t = i\epsilon \Theta,
\end{eqnarray}
where $\Theta$ and $\epsilon$ are real Grassmannian paramenters,

\begin{equation}
\label{E3}
[\Theta , \epsilon ]_+ =\Theta\epsilon + \epsilon\Theta = 0 \Rightarrow
(\Theta \epsilon )^{*}= (\epsilon^{*}\Theta^{*}) = (\epsilon\Theta) =
-(\Theta \epsilon).
\end{equation}
This star operation of a product of two anticommuting Grassmannian variables
ensures  that the product is a pure imaginary number
and for this reason must insert the $i=\sqrt{-1}$ in (\ref{E2}) to
obtain the real character of time. SUSY is implemented for
maintain the line element invariante\footnote{That properties of
Grassmannian magnitudes,
necessary for one better comprehension of this section will be introduced
gradatively.}:

\begin{equation}
\label{E4}
dt + i\Theta d\Theta =\hbox {invariant,}
\end{equation}
where one again we introduce an $i$ for the line element to become real.

The supercoordinate for $N=1$ is expanded in a Taylor series in terms
of even $q(t)$ and odd $\psi (t)$ coordinates:

\begin{equation}  \label{E5}
\phi\equiv\phi (t;\Theta ) = q(t) + i\Theta \psi (t).
\end{equation}

Now, we need to define the derivative rule with respect to one
Grassmannian variable. Here, we use the right derivative rule i.e.
considering $f(\Theta_1,\Theta_2)$ a function of two anticommuting variables,
the right derivative rule is the following:

\begin{equation}
\delta f = \frac{\partial f}{\partial \Theta_1} \delta \Theta_1 +
\frac{\partial f}{\partial \Theta_2} \delta \Theta_2,
\end{equation}
where $\delta \Theta_1$ and $\delta \Theta_2$ appear on the right side of
the partial derivatives.

One infinitesimal transformation of supercoordinate that  obbey the SUSY transformation law given by (\ref{E2}) results in:

\begin{equation}
\label{E6}
\delta \phi (t;\Theta ) = \phi (t^{\prime};\Theta^{\prime}) - \phi (t;\Theta
) = (\partial_t\phi)\delta t+(\partial_{\Theta}\phi)\delta \Theta =
i\epsilon \Theta \dot {q}(t) - i\epsilon \psi (t).
\end{equation}

On the other hand, making an
infinitesimal variation of (\ref{E5}) i.e.
$\delta \phi(t;\Theta ) = \delta q(t) + i\Theta \delta \psi (t)$
and comparing with (\ref{E7}) we obtain the following SUSY transformation
law for the components of the supercoordinate:

\begin{equation}
\label{E7}
\delta q(t) = i\epsilon \psi (t),\qquad \delta \psi (t) = -\epsilon\dot{q}(t).
\end{equation}

Therefore making a variation in the even component we
obtain the odd component and vice versa i.e. SUSY mixes the even and odd
coordinates.
Note from (\ref{E5}) and (\ref{E6}), that the infinitesimal SUSY
transformation law can be
written in terms of the supercoordinate $\phi (t;\Theta )$:

\begin{equation}
\label{E8}
\delta \phi (t;\Theta ) = \epsilon Q\phi (t;\Theta ),\qquad Q =
-\partial_{\Theta } + i\Theta \partial_t,
\end{equation}
 where
$\partial_{\Theta}
\equiv\frac{\partial }{\partial \Theta }, \quad\partial_t
\equiv\frac{\partial} {\partial t}.$
Therefore any coordinate which obbey equation (\ref{E8}) will be
interpreted as supercoordinate\footnote{Note that $\delta \tau(t;\Theta)
 = \epsilon Q \tau(t;\Theta)$ (where $\tau$ is a supercoordinate) give us
a way for test if $\tau$ is a supercoordinate indeed. If this equality not
is true, $\tau$ is not a supercoordinate.}. The differential operator $Q$,
called the supercharge, is a representation of the translation generator in the
superspace. In fact one finite translation can be easily
obtained (\ref{E8}) which has an analogous form as that of
translation in the ordinary space

\begin{equation}
\label{E9}
U(\epsilon ) \phi (t;\Theta ) U^{-1}(\epsilon ) = \phi
(t^{\prime};\Theta^{\prime}),\qquad U(\epsilon )=\exp (\epsilon Q),\quad
U^{-1}(\epsilon ) = U(-\epsilon ),
\end{equation}
with the operator $Q$ doing a similar role as that of the linear momentum
operator in ordinary space.

\section{COVARIANT DERIVATIVE AND THE LAGRANGIAN}
\label{sec: level3}

Nowm, we build up a covariant derivative (with respect to $\Theta$) which
preserves the supersymmetry of super-action i.e. we will see that
the derivative with respect to $\Theta$ ($\partial_{\Theta} \Phi)$ does not
itself transform like a supercoordinate. So it is necessary to construct a
covariant derivative.

SUSY really possesses a peculiar characteristic. As the anticommuting
parameter
$\epsilon$ is a constant we see that SUSY is a global symmetry.
In general local symmetries are the ones which require covariant derivatives.
For
example the gauge theory $U(1)$ with local
symmetry requires covariant derivatives. But because of the fact that
$\partial_{\Theta }\phi (t;\Theta )$ is not a supercoordinate, SUSY will
require a covariant derivative for us to write the super-action in a
consistent form. To prove this fact we use (\ref{E9}) so as to obtain the
following variations:

\begin{eqnarray}
\label{E10}
\delta \partial_{\Theta }\phi (t;\Theta ) &&= \partial_{\Theta }
\delta\phi(t;\Theta ) = - i\epsilon \dot{\phi }(t;\Theta ) + i\epsilon \Theta
\partial_{\Theta }\dot{\phi }(t;\Theta ) \neq
\epsilon Q\partial_{\Theta}\phi (t;\Theta ) \nonumber\\
\delta \partial_\Theta \phi(t;\Theta) &&= \epsilon\Theta\dot{\psi }(t) \neq
\epsilon Q \partial_\Theta\phi(t;\Theta) = -\epsilon\Theta\dot{\psi }.
\end{eqnarray}
On the other hand, making an infinitesimal variation of the partial temporal
derivative we find:

\begin{equation}  \label{E11}
\delta \partial_{t}\phi (t;\Theta ) = \epsilon Q\partial_{t}\phi (t;\Theta ).
\end{equation}
So we conclude that $\partial_{t}\phi $ obeys the SUSY transformation law
and therefore it is a supercoordinate. The covariant derivative of
supersymmetric classical mechanics is constructed so that it obeys the
anticommutativity with $Q$ i.e. $[D_{\Theta }, Q]_{+}= 0.$ It is easy to
verify that one
representation for a covariant derivative is given by:

\begin{equation}
\label{E12}
D_{\Theta } = -\partial_{\Theta } - i\Theta \partial_{t} \Leftrightarrow
\delta D_{\Theta }\phi (t;\Theta ) = \epsilon QD_{\Theta }\phi (t;\Theta ).
\end{equation}
Another interesting property which occures when the SUSY generator
($Q$) is realized in terms of spatial coordinates on in the configuration representation
\cite{Ber84,Gen} is the fact that the anti-commutator of the operator
$Q$ with itself gives us the SUSY Hamiltonian:

\begin{equation}
\label{E13}
[Q, Q]_{+} =-2i\partial_{t} = -2H, \quad Q^2 =-H, \quad (SUSI)^{2} \propto H,
\end{equation}
i.e. two sucessive SUSY transformations give us the Hamiltonian.
This is an algebra of left supertranlations and time-translations.
The corresponding right-supertranslations satisfy the following algebra:

\begin{equation}
\label{ED}
[D_{\Theta}, D_{\Theta}]_{+} =2i\partial_{t} = 2H, \quad D_{\Theta}^2 =H.
\end{equation}

Before we construct the Lagrangian for a superpoint participle, we introduce the
Berezin integrals\cite{Berezin66} for an anticommuting variable:

\begin{equation}  \label{E14}
\int d\Theta\Theta = 1 = \partial _{\Theta }\Theta ,\qquad \int d\Theta = 0
= \partial _{\Theta }1.
\end{equation}

Now we are in conditions to analyse the free superpoint particle in one
dimension and to construct a manifestly supersymmetric action.
We will see that SUSY is a super-action symmetry but does not
let the Lagrangian invariant. A super-action for the free superpoint particle
can be written as the following double integral\footnote{In this section about
supersymmetry we use the unit system in which $m=1=\omega$, where $m$ is
the particle mass and $\omega$ is the angular frequency.}

\begin{eqnarray}
\label{E15}
S &=& {\frac{i}{2}}\int\int {dtd}\Theta (D_{\Theta }\phi )
\dot{\phi } \nonumber\\
  &=& {\frac{i}{2}}\int\int {dtd}\Theta \{-i\psi \dot{q} - \Theta \psi
 \dot{\psi } - i\Theta\dot {q}^{2}\} \nonumber\\
  &=& -\frac{i}{2} \int {dt} \{i\psi\dot{q} \int {d\Theta} + \psi\dot{\psi}
 \int {d\Theta\Theta} + i\dot{q}^2 \int {d\theta\Theta} \} \nonumber\\
  &\equiv& \int {dtL}.
\end{eqnarray}

Indeed after integrating in the variable $\Theta $, we obtain the following
Lagrangian for the superpoint particle:

\begin{equation}
\label{E16}
L = {\frac{1}{2}} \dot{q}^{2}- {\frac{i}{2}}\psi \dot\psi,
\end{equation}
where the first term is the kinetic energy associated with the even
coordinate in which the mass of the particle is unity. The second term is a
kinetic energy piece associated with the
odd coordinate (particle's Grassmannian degree of freedom) dictated by
SUSY and is new for a particle without potential energy. Thus we see that the
Lagrangian is not invariant because it's variation result in a total
derivative and consequently is not zero, which can be obtained from
$\delta S,\, D_{\Theta }\mid_{\Theta =0}= Q_{\Theta }\mid_{\Theta =0} $:

\begin{equation}
\label{E17}
\delta S = {\frac{i}{2}}\int {dtd}\Theta \delta \{(D_{\Theta }\phi )
\dot{\phi }\} \Rightarrow\delta L = {\frac{1}{2}} \epsilon {\frac{d}{dt}}
(D_{\Theta }\phi )\dot{\phi } \mid_{\Theta =0} = {\frac{i}{2}}
\epsilon {\frac{d}{dt}}\{\psi \dot{q}\}\neq 0.
\end{equation}

Because of the fact that the Lagrangian is a total derivative, we obtain
$\delta S = 0$ i.e. the super-action is invariant under N=1 SUSY
transformation.

Note that for $N=1$ SUSY  and with only one coordinate $\phi$, we can\'{}t
introduce a potencial term $ V(\phi)$ in the super-action because
it conduces to non-invariance i.e. ($\delta S \neq 0$). There are even two
more inconsistency problems. First we note that the super-action $S$ acts
like an even element of the Grassmann algebra and for this reason any
additional
term must be an even element of this algebra. Indeed, analysing the terms
present in the super-action we see that the line element has one
$d\Theta$ and one $dt$ which are respectively odd and even. As the
supercoordinate is even, the potential $V(\phi)$ must also be even, which
when acts with the line element $dtd\Theta$ becomes odd which fact
will let the super-action odd and this is not admissible. The other
inconsistency problem can be traced form the dimensional analysis. In
the system of natural units the super-action must be non-dimensional. In such a system of units, the time and the even component $q(t)$ of the supercoordinate have
dimension of $[massa]^{-1}$. In this way, starting form the supertranslation,
we will see that $\Theta$ will have dimension $[massa]^{-\frac{1}{2}}$.
Consequently the supercoordinate $\phi$ has dimension of $[massa]^{-1}$
and $\dot\phi$ is non-dimensional. Because of this when we
introduse a potential term
$V(\phi)$ we would obtain a super-action with inconsistent dimension.

The cannonical conjugate momentum associated to the
supercoordinate is given by

\begin{equation}
\label{E18}
\Pi(t,\Theta)= \frac{\partial}{\partial\frac{d\phi}{dt}} L=
\frac{i}{2}\frac{\partial}{\partial\frac{d\phi}{dt}}\int d
\Theta (D_{\Theta }\phi )\dot{\phi } =
\frac{1}{2}\int d\Theta \{\Theta \dot{\phi }+iD_{\Theta }\phi \},
\qquad \dot{\phi }\equiv\frac{d\phi}{dt},
\end{equation}
which leads to the following  Poisson brackets:

\begin{eqnarray}
\label{E19}
\{\phi(t,\Theta), \phi(t,\Theta^{\prime})\} =0= \{\Pi(t,\Theta),
\Pi(t,\Theta^{\prime})\} \\
\{\phi(t,\Theta), \Pi(t,\Theta^{\prime})\}=\delta(\Theta-\Theta^{\prime}).
\end{eqnarray}

We can not implement the first cannonical quantizating method directly
because there exist constraints: the primary obtained from the definition of cannonical
momentum and the secondary obtained from the consistency condition.
In this case we must to construct the modified Poisson parentheses
called Dirac brackets. These aspects have been considered in the
quantization of the superpoint particle with extended $N=2$ SUSY and is
out of the scope of this work \cite{Barce}.

We finalize by writing another manifestly supersymetric action
which can be constructed for the case with $N=1$ SUSY
using the generator of right supertranslation $D_{\Theta}$:

\begin{equation}
\label{S2}
S_2=\frac{i}{2}\int\int d\Theta dtD_{\Theta}(D_{\Theta}\phi).
\end{equation}

It is left as an exercise for the reader to demonstrate that it is
possible to effect the integral in $\Theta$ and encounter the
$N=1$ SUSY  Lagrangian for the case.

\section{ CONCLUSION \protect\\}
\label{sec:level4}

After the introduction of a real Grassmannian anticommuting variable,
we consider a translation in  superspace and implement the
transformation laws of the supercoordinate and  its components.
We show that an infinitesimal variation of the even coordinate
generates the odd coordinate and vice-versa, characterizing N=1 SUSY.
We introduce a covariant derivative for writing the super-action
in a consistent way. We verify that occure an interesting property
occurs when the SUSY generator ($Q$) is realized in terms of
Grassmannian coordinates: the
anticommutator of $Q$ with itself results in the Hamiltonian i.e.
two successive SUSY transformations gererate the Hamiltonian.
If the reader considers two successive supertranslations, will
obtain exactly the Hamiltonian as result i.e. $D^2_{\Theta}= H$.
In the original works
about supersymmetric in classical mechanics \cite{Galv80,Salo82}, the
respective authors do not justify as to why in the case of N=1 SUSY
is not allowed to put a potential term in the Lagrangian.
Therefore, the main
purpose in this work has been to make an analysis of this question
in the context of a Lagrangian formalism in superspace with N=1 SUSY.
In synthesis from the fact as to how the super-action must be even
and the line
element $dtd\theta$ in its construction be odd, we show that it is not
possible to introduction a potential energy term $V(\phi)$, because
which a potential term would conduze in a super-action with inconsistent
dimension i.e. the super-action itself becomes odd too. Therefore
when we have only one supercoordinate $\phi$, the $N=1$ SUSY exists
only for a free superpoint particle. The equations of motion for the
superpoint particle with N=1 SUSY are first order for odd coordinate
$(\dot\Theta= \frac{d\Theta}{dt}=0)$ and second order for even
coordinate $(\ddot x= \frac{d^2}{dt^2}x=0)$.

In conclusion we must stress that the super-action must always be even but
the Lagrangian may eventually  be odd.
Nonetheless, the same analysis can be implemented for the case
with $N=2$ SUSY so that one may put a potential term in the superaction.
In this case, considering only one
supercoordinate $\phi$ of commuting nature, is allowed the introduction
of a potential term in the Lagrangian \cite{Barce,Barce97}.
 On the other hand, one can introduce an
odd supercoordinate of anticommuting nature
($\Psi(\Theta;t)=\psi(t)+q(t)\Theta)$ so that the  $N=1$ SUSY
is ensured and the main consequence is to obtain the
unarmonic oscillator potential.

\vspace{0.5cm}

\centerline {\bf {ACKNOWLEDGMENTS}}

\vspace{0.5cm}

The authors are grateful to the Departamento de F\'{\i}sica da Universidade
Federal da Para\'{\i}ba, Campus II, and CNPq for support.
WPA and IFN are indebted
the {\it Conselho Nacional de Desenvolvimento Cient\'{\i}fico e
Tecnol\'{o}gico (CNPq)} for partial financial support
trough the PIBIC/UFPB/CNPq program.
Thanks are also due to J. A.
Helayel Neto for hospitality at CBPF-MCT and for fruitful discussions on
supersymmetric models.



\end{document}